\documentstyle[12pt]{article}
\makeatletter
\@addtoreset{equation}{section}

\makeatother
\def\G1{\displaystyle \mathop{G}}
\def\e1{\displaystyle \mathop{e}}
\def\u1{\displaystyle \mathop{u}}

\def\n1{\displaystyle \mathop{\nu}}
\def\ps1{\displaystyle \mathop{\Psi}}

\def\p1{\displaystyle \mathop{p}}
\def\J1{\displaystyle \mathop{J}}

\def\O1{\displaystyle \mathop{O}}

\def\hp1{\displaystyle \mathop{\hat{p}}}

\def\FFr{\displaystyle \frac}

\def\sig1{\displaystyle \mathop{\sigma}}

\def\h1{\displaystyle \mathop{h}}
\def\H1{\displaystyle \mathop{H}}
\def\A1{\displaystyle \mathop{A}}

\def\D1{\displaystyle \mathop{D}}
\def\D1{\displaystyle \mathop{D}}
\def\g1{\displaystyle \mathop{g}}
\newtheorem{guess}{Conjecture}

\begin{document}

\begin{titlepage}
\begin{center}

\null
\vskip-1truecm
\rightline{IC/94/290}
\vskip1truecm
{International Atomic Energy Agency\\
and\\
United Nations Educational Scientific and Cultural Organization\\
\medskip
INTERNATIONAL CENTRE FOR THEORETICAL PHYSICS\\}
\vskip3truecm
{\bf INTRODUCTION TO THE THEORY OF GOYAKS\\ \it OPERATOR MANIFOLD
APPROACH \\ TO GEOMETRY AND PARTICLE PHYSICS\\}
\vskip3truecm
{G.T.Ter-Kazarian \footnote{\normalsize Permanent Address:
Byurakan Astrophysical Observatory, Armenia 378433.\\
E-mail address:byur@mav.yerphi.am }\\

International Centre for Theoretical Physics, Trieste, Italy,\\}
\end{center}
\vskip1truecm

\vskip3truecm

\begin{center}
{MIRAMARE -- TRIESTE\\
\medskip
September 1994\\}
\end{center}

\end{titlepage}

\centerline{ABSTRACT}
\baselineskip=24pt
\bigskip
The fundamental question that guides our discussion is "how did the
geometry and particles come into being?" To explore this query
we suggest the theory of goyaks~\footnote{\normalsize{\em a goyak
in Armenian means an existence (an existing structure). This term has been
firstly used in (Ter-Kazarian, 1986)}}, which reveals the
primordial deeper structures underlying
fundamental concepts of contemporary physics. It address itself to the
question of the prime-cause of origin of geometry and basic concepts
of particle physics such as the fundamental fields of quarks and leptons
with the spins and various quantum numbers, internal symmetries and so
forth; also
basic principles of Relativity, Quantum, Gauge and Color Confinement,
which are, as it was proven, all derivative and  come into being
simultaneously. The substance out of which the geometry and particles are
made is a set of new physical structures - the goyaks, which are involved
into reciprocal linkage establishing processes.

The most promising aspect of our approach so far is the fact  that
many of the important anticipated properties, basic concepts and
principles of particle physics are appeared quite naturally in the framework
of suggested theory.

In pursuing the original problem further
we have elaborated a new mathematical framework in order to describe the
persistent processes of creation and annihilation of goyaks in the definite
states. It is, in fact, a still wider generalization of familiar
methods of secondary
quantization with appropriate expansion over the geometric objects.

One interesting offshoot of this generalization directly leads to
the formalism of operator manifold $\hat{G}(2.2.3)$, which will
frame our discussion throughout this paper.
In broad sense it consequently yields the quantization of geometry, which
differs in principle from all earlier studies.
It was found out a contingency  arisen at the very beginning that all
states of goyaks are degenerate with the degree of degeneracy equal 2.
That is, the goyaks are turned out to be fermions with half-integral spins,
which subsequently give rise to origin of the spins of particles.

The nature of $\hat{G}(2.2.3)$ provides its elements with both the field
and geometric aspects.

We have directly passed from $\hat{G}(2.2.3)$ to wave-manifold
$\widetilde{G}(2.2.3)$, which is practically made of goyak's
eigen-functions and was endowed by a statistical-probabilistic
nature.

We have dealt with the principles of operator $\hat{F}$ and
wave $\widetilde{F}$ groups and their representations, which
refer to continuous and discrete symmetries of operator and
wave manifolds.

To render the discussion more transparent the generalized
causal Green's functions of regular goyaks in
terms of wave manifold have been introduced, by means of which
the geometry realization condition has been cleared up. There are
two different choices of realization of geometry. First one is a
trivial, which leads to geometry without particles and interactions.
But the second choice subsequently yields the geometry G(2.2.3) with
the particles and interactions.

We employ the wave functions of distorted goyaks to extend the
knowledge here gained regarding the quark fields are being introduced
in the color space of internal degrees of freedom. The local distortion
rotations around each fixed axis yield the quarks or anti-quarks.
They obey exact color confinement, and the spectrum of hadrons would emerge
as the spectrum of the color singlet states.

Finally, it has been shown that the gauge principle holds for any
physical system which can be treated as a definite system of distorted goyaks.

The theory of goyaks predicts a class of possible models of internal
symmetries, which utilize the idea of gauge symmetry and reproduce the
known phenomenology of electromagnetic, weak and strong interactions.

Here we focused our attention mainly on developing the mathematical
foundations for our viewpoint. Hence our discussion has been rather
general and abstract. Of course, much remains to be done for a larger
contribution into the particle physics.
However, we believe that the more realistic final theory of particles and
interactions can be found within a context of suggested theory of goyaks.

\vskip1truecm

\begin{newpage}

\section {Introduction}
\label {int}

The contemporary understanding of physical processes is mainly based
on the concepts of local symmetries and gauge fields (Weil, 1918;
Yang et al, 1954;
Utiyama, 1956; Glashow, 1961; Salam, 1959; Schwinger, 1962; Kibble, 1961).

Non-Abelian gauge theories may have something to do with nature. They are
utilized by theoreticians for the description of both electro-weak
interactions (Salam et al, 1964; Schwinger,1957; Glashow,1961; Weinberg, 1967;
Salam, 1968; Abers et al, 1973) and strong interactions (Fritzsh et al, 1972;
Weinberg, 1973; Gross et al, 1973; Marciano et al, 1978).

The theoretical picture of the weak interactions has became rather definite.
At present a consistent description of the known weak interactions can
be given in terms of gauge theory based on the group
$SU^{loc}(2)\otimes U^{loc}(1)$.

Many physicists suspect that the underlying theory of strong interactions
is the gauge theory with the local unitary group $SU^{loc}_{C}(3)$,
consisting of colored quarks of various types coupled to eight colored
gauge bosons (gluons).
The existence of the internal symmetry group  $SU^{loc}_{C}(3)$ allows
oneself to introduce a gauge theory in color space, with the color charges
acting as exactly conserved quantities.
The role played by quarks in hadron physics has became
rather transparent. In building up the particle spectrum, they act as
constituents such that states of the type $qqq$ describe the baryons, and
states of the type $q\bar{q}$ the mesons. One can postulate the confinement
condition such that all physical particles and observables are required
to be singlets under the local color transformations  implemented on the
quarks through the  $SU^{loc}_{C}(3)$
rotation matrices in the fundamental representation. This leads automatically
to the correct spectrum and eliminates quarks $(q)$, diquarks $(qq)$ etc. as
physical particles. That is, a color would be confined and the spectrum of
hadrons would emerg as the spectrum of the color singlet states. This
indicates to the hidden realization of a symmetry: only singlets of the
symmetry group exist as the physical states. This possibility requires the
symmetry to be an exact symmetry.

At present there is a strong tendency among theoreticians to believe in the
forceful arguments brought up in favor of absolute confinement. It remains
to be seen if there is the absolute color confinement, which is certainly
the simplest and most attractive possibility, but perhaps not the one chosen
by nature.

It will be advantageous to unify all interactions within one simple
group. Several models of this type have been discussed by various authors
(Weinberg, 1972; Pati et al, 1973; Georgi \& Glashow, 1974;
Georgi et al, 1974, 1979; Gunaydin et al, 1974; Ramond, 1977).
As one might expected from a theory, which unifies
the color and flavor groups of the quarks and the flavor group of the
leptons in one simple group, the conservation of baryon number or the
stability of the proton with respect to its decay into leptons is a
nontrivial aspect (Gell-Mann et al, 1978; Harari, 1978; Goldman et al, 1980;
Ellis et al, 1980; Segre et al, 1980). Which of the various schemes,
if any, is realized either exactly or at least approximately in nature,
remains to be seen.

The intensive attempts have been made in order to construct the unified
field theory of particles and all interactions including also the
gravitation, based on the concepts of supersymmetries and superstrings
(Wess et al, 1974; Freedman et al, 1976; Deser et al, 1976; Schwarz,1984;
van Nieuwenhuizen, 1981; Schwarz, 1985; Green et al, 1985).
There are a number of advantages to this suggestion,
but on the theoretical side there are also several serious
problems one has to deal with in these schemes.
Of course, much remains to be done before one can determine whether this
approach can ever make a larger contribution into the physics of
particles and interactions.

Although a definite pattern for the theoretical description of particle
physics has emerged, which is attractive enough both theoretically and
phenomenologically, but it could not be regarded as the final word in
particle physics, and many fundamental questions have yet to
be answered. It is worth emphasizing that it is yet very vulnerable,
especially on the theoretical basis.
As a last remark, we note that just as in these earlier studies,
the fundamental concepts and right
symmetries, also basic principles of particle physics have been put
into theory by hand. The difficulties associated with this step are
notorious.

Thus, in spite of the considerable progress achieved over the
entire subsequent period in the study of the fundamental constituents
of the matter and forces, the physical theory is still far from being
complete, and not free of many serious difficulties. Moreover, we are
only at the very beginning of deeper understanding of the prime-cause
of the origin of the basic concepts and principles of the particle physics,
which are uncertain by now and there is still a long way to go.

This brings us to the greatest questions of physics: How did the fundamental
concepts such as the space-time continuum, quarks and leptons, their
masses at rest, various quantum numbers; and also basic principles of physics
such as Relativity, Quantum, Gauge and Color Confinement principles come
into being? Are they all primordial or derivative?

To explore these queries we are led to consider
the big problem, as it was seen at the outset, which is how to find out
the substance or basic fundamental structures, something deeper
than the geometric continuum
of four dimension (representing the arena of space-time, within which
the phenomenon
of the particles is presumed to appear), that underly both geometry and
particles?

It is likely that these questions cannot be answered within the scopes of
conventional physics. The absence of the vital physical theory, which is able
to reveal these structures and answer to the right questions mentionted above,
imperatively stimulates the search for general constructive principles, which
becomes of paramount importance for particle physics.

Following Ter-Kazarian (1986, 1989, 1992) we learned that the manifold
$G(2.2.3)$ underlies the space-time continuum. The perception of the space and
time has been suggested by means of new concepts. While the problems and
dynamics of the processes, which are of interest of Relativity and Quantum
Field Theories, have been studied from a specific novel point of view.
The spacial-time concepts have been properly substituted by the appropriate
new ones.

On the examples of simple fields the possibility of introducing a concept of
mass at rest of particle as a function of the internal degree of freedom has
been shown.

The general theory of distortion of space-time has been suggested. Within
its scopes, as it is widely believed, the space-time is not pre-determined
background on which physical processes take place, but a dynamical entity
itself.

Curvature of space-time continuum is considered as a particular regime
of distortion, which enables one to construct a new theory of gravitation.

The other regime of "pure-inner distortion" of the space-time continuum
below some small length scale
has been discussed. It is shown that the theory predicts new physical
phenomena, which have a dominant contribution in high energy region
with respect to all other processes.

This theory in practical application in astrophysical problem, for example,
enables oneself to provide an alternative approach for understanding of
the internal structure of suppermassive compact stationary celestial bodies
such as active galactic nuclei.
There are a number of advantages to this
approach, which differs in principle from the
standard black hole accretion models (Ter-Kazarian, 1989, 1989, 1990, 1991,
1992) and is in good agreement with the observational data.

The guiding line framing our discussion throughout this
paper is a generalization and further
expansion of basic ideas of the theory of distortion of space-time continuum,
in order to find out the deeper structures underlaying both geometry $G(2.2.3)$
and particles. It will be appropriate to turn to them as the primordial deeper
structures. Suggested theory will address itself to the question of the origin
of fundamental concepts of particle physics such as the quarks, leptons, their
spins, internal symmetries and associated with them different charges. Within
its context we will clear up the prime-cause of origin of basic principles,
especially of the most challenging ones such as Gauge and Color Confinement
principles.

The paper is organized as follows: Being confronted by the problems
mentioned above, in first part of treatment our task will be to develop
and understand the conceptual foundations for our viewpoint in general.
To begin with a
description of theory we choose a simple setting and consider new formations
designed to endow certain physical properties and satisfying the rules
stated in section 2. It forms the starting point for them to get into
the processes of
establishing the reciprocal "linkage" between different type of
goyaks in given state.

We will elaborate a new mathematical framework in order to describe the
persistent processes of creation and annihilation of goyaks. It will be,
in fact, a still wider generalization of the familiar methods of secondary
quantization
or similar processes, with appropriate expansion over the geometric objects.
Hence it gives rise to the formalism of "operator manifold" $\hat{G}(2.2.3)$.
We will discuss two aspects of it. First one is a quantum field aspect, but
a second- is a differential geometric aspect.

A closer examination of the properties of this formalism practically
compulsory leads to the other formalism of "wave manifold"
$\widetilde{G}(2.2.3)$. The latter is a manifold of eigen-functions of
goyaks and endows a statistical-probabilistic nature.

This one, in its turn, precedes to ordinary geometry $G(2.2.3)$, within
which subsequently the phenomena of fundamental fields with the
half-integral spins, different charges and the unitary groups of
internal symmetries emerge.

In the second part, which is devoted mainly to the dynamics of
distorted goyaks, we will formulate a principle of identity of
regular goyaks and extend the knowledge gained in outlined
mathematical framework regarding the origin of
fundamental fields, Gauge and Quark
Confinement principles, also internal symmetries.

In last section we will discuss a class of models of internal
symmetries, which reproduce the known phenomenology of
electromagnetic, weak and strong interactions.
In order to save writing we guess it worthwhile to leave the other
concepts such as the flavors and so forth with associated aspects
of particle physics for an other treatment. It will not concern us
here and must be further discussed. Surely this is an important
subject for separate research.

Here, as far as we should fix our attantion mainly on developing
the mathematical foundations for our viewpoint, a discussion would be
rather general and abstract.
Actually, this kind of mathematical treatment will necessarily
be schematic and introductory by nature, since a complete
discussion requires more realistic investigation.

However, we believe we would make good headway by presenting a reasonable
framework whereby one will be able to verify the basic ideas and illustrate
the main features of the theory of goyaks.

\end{newpage}
\begin {center}{\bf\huge Part I. Regular Goyaks}
\end {center}
\section {The Goyaks and Link-Establishing Processes \\Between Them}
\label {goy}
Before embarking on a description of goyaks, just a very brief recapitulation
of the main points of the structure of manifold (Ter-Kazarian, 1989, 1992)
\begin {equation}
\label {R21}
G(2.2.3)={}^{*}G(2.2)\otimes G(3) ,
\end{equation}
with the set of unit elements $\{e_{(\lambda,\mu,\alpha)}\} \quad
(\lambda,\mu=1,2;
\quad \alpha=1,2,3)$ providing the basis
\begin{equation}
\label {R22}
e_{(\lambda,\mu,\alpha)}=O_{\lambda,\mu}\otimes\sigma_{\alpha} .
\end{equation}
The set of the linear unit bi-pseudo-vectors $\{ O_{\lambda,\mu} \}$
\begin{equation}
\label {R23}
<O_{\lambda,\mu},O_{\tau,\nu}>={}^{*}\delta_{\lambda,\tau}
{}^{*}\delta_{\mu,\nu},
\quad {}^{*}\delta_{\lambda,\tau}=\cases{1 \quad \rm if\quad
{\lambda \neq \tau},\cr 0  \qquad \rm otherwise,\cr}
\end{equation}
is the basis in $2 \times 2$ dimensional linear bi-pseudo-space
${}^{*}G(2.2).$ \\
The $G(3)$ is the three-dimensional real linear space with the
basis consisted of the ordinary unit vectors $ \sigma_{\alpha}$
\begin{equation}
\label {R24}
<\sigma_{\alpha}, \sigma_{\beta}>= \delta_{\alpha\beta}=\cases{1
\quad \rm if\quad {\alpha = \beta},\cr 0  \qquad \rm otherwise.\cr}
\end{equation}
Henceforth we always let the first two subscripts in the parentheses
specify the bi-pseudo-vector components, the third refers to the
ordinary-vector components.\\
The metric in $G(2.2.3)$ is bilinear, local, symmetric and positive
defined reflection
\begin{equation}
\label {R25}
\hat{g}:T_{p}\otimes T_{p}\rightarrow C^{\infty}(G(2.2.3)),
\end{equation}
where $T_{p}$ is the whole set of vector fields on $G(2.2.3)$.\\
The metric $\hat{g}$ reads in component form with respect to the
basis $\{ e_{(\lambda,\mu,\alpha)} \}$
\begin{equation}
\label {R26}
g_{(\lambda,\mu,\alpha)(\tau,\nu,\beta)}=g(e_{(\lambda,\mu,\alpha)},
e_{(\tau,\nu,\beta)})=g(e_{(\tau,\nu,\beta)},e_{(\lambda,\mu,\alpha)}).
\end{equation}
Bilinear form on vectors $\zeta=e^{(\lambda,\mu,\alpha)}\zeta_
{(\lambda,\mu,\alpha)} \in G(2.2.3)$ with respect to holonomic
basis is given as follows:
\begin{equation}
\label {R27}
\hat{g}=g^{(\lambda,\mu,\alpha)(\tau,\nu,\beta)}d\zeta_
{(\lambda,\mu,\alpha)}\otimes d\zeta_{(\tau,\nu,\beta)}.
\end{equation}
Analogical form on co-vectors  $\bar\zeta=e_{(\lambda,\mu,\alpha)}\zeta^
{(\lambda,\mu,\alpha)}$ reads
\begin{equation}
\label {R28}
\hat{g}=g_{(\lambda,\mu,\alpha)(\tau,\nu,\beta)}d\zeta^
{(\lambda,\mu,\alpha)}\otimes d\zeta^{(\tau,\nu,\beta)}.
\end{equation}
Except where stated otherwise, here as usual,
the double occurrence of the dummy indices will be taken
to denote a summation extended over their all values.\\
The manifold $G(2.2.3)$ decomposes into two 6-dimensional manifolds
\begin{equation}
\label {R29}
G(2.2.3)=\G1_{\eta}(2.3)\oplus \G1_{u}(2.3).
\end{equation}
The set of unit elements
\begin{equation}
\label {R210}
{\e1_{i}}^{0}_{(\lambda \alpha)}= {\O1_{i}}_{\lambda} \otimes \sigma_{\alpha},
\quad (\lambda=\pm; \alpha=1,2,3)
\end{equation}
is the basis in manifolds $\G1_{\eta} (2.3)(i=\eta)$ and  $\G1_{u}(2.3)(i=u)$,
in which the positive metric forms are defined
\begin{equation}
\label {R211}
\begin{array}{l}
\eta^{2}=\eta_{(\lambda\alpha)}\eta^{(\lambda\alpha)}
\in \G1_{\eta}(2.3), \\
u^{2}=u_{(\lambda\alpha)} u^{(\lambda\alpha)}\in \G1_{u}(2.3),
\end{array}
\end{equation}
where
\begin{equation}
\label {R212}
\begin{array}{l}
{\e1_{i}}^{0}_{(+ \alpha)}= \FFr{1}{\sqrt{2}}(e_{(1,1,\alpha)}+
\varepsilon_{i} e_{(2,1,\alpha)}),\\
\\
{\e1_{i}}^{0}_{(- \alpha)}= \FFr{1}{\sqrt{2}}(e_{(1,2,\alpha)}+
\varepsilon_{i} e_{(2,2,\alpha)}),\\
\\
\varepsilon_{i}=\cases{\quad1 \quad \rm if\quad
{i = \eta},\cr \quad\hspace{-.25cm}-1  \quad \rm if \quad {i=u},\cr}
\end{array}
\end{equation}
and
\begin{equation}
\label {R213}
\begin{array}{lr}
\eta_{(+ \alpha)}= \FFr{1}{\sqrt{2}}(\zeta_{(1,1,\alpha)}+
\zeta_{(2,1,\alpha)}),   \qquad u_{(+ \alpha)}= \FFr{1}{\sqrt{2}}(\zeta_
{(1,1,\alpha)}-\zeta_{(2,1,\alpha)}), \\
\\
\eta_{(- \alpha)}= \FFr{1}{\sqrt{2}}(\zeta_{(1,2,\alpha)}+
\zeta_{(2,2,\alpha)}),   \qquad u_{(- \alpha)}= \FFr{1}{\sqrt{2}}(\zeta_
{(1,2,\alpha)}-\zeta_{(2,2,\alpha)}).
\end{array}
\end{equation}
The following scalar products of vectors (co-vectors) are available
\begin{equation}
\label {R214}
\begin{array}{ll}
<{\O1_{i}}_{\lambda},{\O1_{j}}_{\tau}>=\varepsilon_{i}\delta_{ij}
{}^{*}\delta_{\lambda\tau} \qquad (<{\O1_{i}}^{\lambda},{\O1_{j}}^
{\tau}>=\varepsilon_{i}\delta_{ij}{}^{*}\delta^{\lambda\tau}),\\
\\
<{\O1_{i}}^{\lambda},{\O1_{j}}_{\tau}>=\varepsilon_{i}\delta_{ij}
\delta^{\lambda}_{\tau},
\end{array}
\end{equation}
and
\begin{equation}
\label {R215}
<{\e1_{i}}_{(\lambda\alpha)}^{0},{\e1_{j}}_{(\tau\beta)}^{0}>
=\varepsilon_{i}\delta_{ij}
{}^{*}\delta_{\lambda\tau}\delta_{\alpha\beta}, \qquad
<{\e1_{i}}^{(\lambda\alpha)}_{0},{\e1_{j}}_{(\tau\beta)}^{0}>
=\varepsilon_{i}\delta_{ij}
\delta^{\lambda}_{\tau}\delta_{\alpha\beta},
\end{equation}
provided
\begin{equation}
\label {R216}
{\O1_{i}}^{\lambda}={}^{*}\delta^{\lambda\tau}{\O1_{i}}_{\tau},\qquad
{\e1_{i}}^{(\lambda\alpha)}_{0}={}^{*}\delta^{\lambda\tau}
\delta^{\alpha\beta}{\e1_{i}}_{(\tau\beta)}^{0}.
\end{equation}
To render our discussion here more transparent
next we will develop the foundations for our viewpoint and
proceed to general definitions and conjectures of the theory of
goyaks directly. At this, for simplest sense, we may consider it
briefly hoping to mitigate a shortage of insufficient rigorous treatment
by the further exposition of the theory and make them complete and discussed
in broad sense in due course.

\begin{guess}
The $6$-dimensional basis vectors ${\e1_{i}}^{0}_{(\lambda \alpha)}$
(or co-vectors ${\e1_{i}}_{0}^{(\lambda \alpha)}$)
we explore from a specific novel point of view, as being
the main characteristics of the real existing structures called "goyaks".
Below we distinguish two type of goyaks: $\eta$-type ($i=\eta$) and $u$-type
$(i=u)$, respectively.
\end{guess}
\begin{guess}
The goyaks establish reciprocal "linkage" between themselves. The
links are described by means of "link-function" vectors
${\ps1_{\eta}} _{(\lambda\alpha)}(\eta,p_{\eta})$ and
${\ps1_{u}}_{(\lambda\alpha)}(u,p_{u})$ (or co-vectors
${\ps1_{\eta}}^{(\lambda\alpha)}(\eta,p_{\eta})$ and
${\ps1_{u}}^{(\lambda\alpha)}(u,p_{u})$):

\begin{equation}
\label {R217}
{\ps1_{\eta} }_{(\pm\alpha)}(\eta,p_{\eta})=\eta_{(\pm\alpha)}
{\ps1_{\eta} }_{\pm}(\eta,p_{\eta}),
\end{equation}
\end{guess}

FILE WAS RECEIVED TRUNCATED AT THIS POINT.
SUBMITTER DOESN'T CARE, NEITHER DO WE.

\end{document}